\documentclass[aps,prx,twocolumn,superscriptaddress,showpacs,longbibliography]{revtex4-1}
\usepackage{amsmath,amssymb,graphicx,subfigure}

\usepackage[utf8]{inputenc}
\usepackage[T1]{fontenc}
\usepackage{xcolor}

\IfFileExists{newtxtext.sty}
   {\usepackage{newtxtext,newtxmath}}
   {\IfFileExists{stix.sty}
      {\usepackage{stix}}
      {\IfFileExists{mathptmx.sty}
      {\usepackage{mathptmx}}{} } }

\usepackage{textcomp}

\usepackage{bm}

\IfFileExists{siunitx.sty}{\usepackage{booktabs,siunitx}}{}

\pdfoutput=1
\usepackage{color}
\definecolor{LinkColor}{rgb}{0.256,0.439,0.588}
\usepackage{hyperref}
\hypersetup{
   colorlinks=true,
   citecolor=LinkColor,
   linkcolor=LinkColor,
   urlcolor=LinkColor
}

\newcommand{\be}{\begin{equation}}
\newcommand{\ee}{\end{equation}}
\newcommand{\bea}{\begin{eqnarray}}
\newcommand{\eea}{\end{eqnarray}}

\usepackage{pifont}

\begin{document}

\title{Diagnosis of topological nodal lines with nontrivial monopole charge in the presence of rotation symmetries}

\author{Heqiu Li}
\affiliation{Department of Physics, University of Michigan, Ann Arbor, MI 48109, USA}
\author{Chen Fang}
\affiliation{Beijing National Laboratory for Condensed Matter Physics and Institute of Physics, Chinese Academy of Sciences, Beijing 100190, China}
\author{Kai Sun}
\affiliation{Department of Physics, University of Michigan, Ann Arbor, MI 48109, USA}

\date{\today}

\begin{abstract}
We discuss a general diagnosis scheme for topological nodal lines protected by time-reversal symmetry and space-inversion symmetry, in the absence of spin-orbit coupling. It is shown that when a shallow band inversion (with a small inverted gap) appears at the $\Gamma$-point in the Brillouin zone, nodal lines are generated along with the inversion. Much information on the types and numbers of these nodal lines can be related to the rotation eigenvalues and inversion eigenvalues of the inverted bands. In addition, we establish a $\mathbb{Z}$-type monopole charge for nodal lines with fourfold or sixfold rotation axis. We point out that the new monopole charge is only well-defined in a fragile sense, that is, when the orbitals near the Fermi energy are restricted to those having the same sign under a twofold rotation. The diagnosis scheme goes beyond the one discussed in Phys. Rev. X 8, 031069 (2018), particularly when there is a band crossing along some high-symmetry line, or when the nodal line has non-unity monopole charge.
\end{abstract}
\pacs{pacs numbers}
\maketitle

\section{Introduction}

Band inversions are considered by many as the source of topological nontrivial properties in materials~\cite{Bernevig2006,Fu2007,Hasan2010rev,Qi2011Rev}.
The inversion here refers to an anomalous ordering of the energy of orbitals with respect to the order in a single atom.
However, on a lattice, the atomic orbitals are no longer valid descriptions of Bloch wave functions, and the bands at high-symmetry points are labelled by the respective irreducible representations.
Therefore, an exact definition of band inversion in crystal is the reverse of the order in the energies of two irreducible representations at some high-symmetry momentum, compared with their ordering in some atomic insulator.
The quantitative relations between band inversion and the induced nontrivial topology has been studied in great detail in Refs.[\onlinecite{Bradlyn2017,Po2017, Song2018quantitativemappings, Khalaf2018,Kruthoff2017}],
so that knowing the symmetry representations at all high-symmetry points helps the diagnosis of potential topology in any nonmagnetic material~\cite{Zhang2019,Vergniory2019,Tang2019}.
In the presence of spin-orbital coupling, sometimes the band inversion brings the system to a topological insulator~\cite{Kane2005,Bernevig2006,Moore2007} or a topological crystalline insulator~\cite{Fu2011,Hsieh2012,Chiu2016},
but in the absence of spin-orbital coupling, after inversion at high-symmetry momenta, there is always some band crossing ``left-over'' being either Weyl points~\cite{Murakami2007,Wan2011,Armitage2018}, or nodal lines~\cite{Burkov2011,Fang:2015aa,Fang2016rev}.
Physically this is because in the presence of spin SU(2) symmetry the bands are in general easier to cross.

Particularly, in the presence of inversion and time-reversal, nodal lines appear at almost all types of band inversions, a case which one of us has studied in Ref.[\onlinecite{SongPRX2018}].
The method in that paper, however, only deals with bands that do \emph{not} have band crossings along high-symmetry lines, so all the nodal lines are away from high-symmetry lines or points.
While this exclusion greatly simplified the discussion, it also failed to capture interesting band crossing structure like a ``cage'' woven around some high-symmetry line,
or the emergence of nodal line having new types of monopole charge, beyond the $\mathbb{Z}_2$-charge in Ref.[\onlinecite{Fang:2015aa}], which we will study in this paper.
Since the creation and annihilation of nodal lines involve at least four bands, we use a four-band tight binding model to study all the new types of nodal structures and their transitions, while generalization beyond the four-band setup will be discussed in the discussion section.

In a four-band model, having two conduction and two valence bands, respectively, we first show that the Hamiltonian can be simplified into one of the two special limits $H_1$ or $H_2$ in a smooth process without introducing new band crossings between conduction and valence bands.
The relatively simple form, and the doubly degenerate spectrum of $H_{1,2}$ make easy the calculation of the band crossing positions and wave functions.
The band inversion is then studied in the simplified Hamiltonian, where all topological nodes are necessarily point nodes. Defining the so called ``parity Chern number'' for the Fermi surface, we are able to classify the types, configurations and parity Chern numbers of the nodal points for each type of band inversion.
In fact, generalizing a discussion in Ref.[\onlinecite{cfang2012}], we can obtain the quantitative relations between the rotation and inversion eigenvalues of the inverted bands and the parity Chern number of the nodal point.

After the non-simplifying terms are added, $H$ becomes non-degenerate, and the nodal points expand naturally into nodal lines.
It can be expected that some quantum numbers of the nodal line, such as the parity Chern number, will be inherited from the former.
To be more specific, a nodal point with parity Chern number 2 after 
non-simplifying terms evolves into either two nodal loops with unity monopole charge each, or into an interesting nodal cage with monopole charge 2, depending on the specific type of band inversion.
A higher-than-unity monopole charge will be defined in due course, under the assumption that near the Fermi energy the orbitals have the same twofold rotation eigenvalue.
This made charge-2 nodal lines one of the fragile topological band crossings~\cite{Pofragile2018,BradlynFragile2019}, in analogy to the recently studied fragile topological gapped states.

In summary, we complete the theory on band inversion induced topological nodal lines in the presence of inversion, time-reversal and spin rotation symmetries, given the physical assumption of small inversion, that is, the inverted gap is small compared with other energy scales.
Under the assumption, the nodal structures form near the point of band inversion, and we propose that the parity and rotation eigenvalues of the inverted bands determine many aspects of the resultant nodal loops: their types, positions and monopole charges.
In this discussion, we include cases where band crossing occurs along high-symmetry lines, and the small band inversion assumption allows us to define fragile high-monopole charge that more comprehensively describe the certain nodal loops, which cannot be understood in Ref.[\onlinecite{SongPRX2018}].

\section{Adiabatic Hamiltonian reduction for four-band systems}
In this section, we show that the Hamiltonian of an arbitrary four-band system can be adiabatically deformed into one of the two special cases without introducing new band crossings between the two valance and two conduction bands. The existence of this adiabatic path allow us to reduce any Hamiltonian into (one of) these two special cases, where topological analysis is greatly simplified and topological indices mostly are fully dictated by band inversions at high symmetry points.

\subsection{$\Gamma$ matrices and the basis for 4-band Hamiltonians}
In general, a $4\times 4$ Hermitian matrices spans a 16-dimensional linear space, and the following 16 matrices can be used as the basis of this linear space:
(1) the identity matrix $I$, (2) four $\Gamma$-matrices $\Gamma_i$ with $i=1$, $2$, $3$, and $4$, (3) six $\Sigma$-matrices
$\Sigma_{i,j}=i [\Gamma_i, \Gamma_j]/2$ and (4) four matrices $i \Gamma_j\Gamma_k\Gamma_l$ where $i=\sqrt{-1}$ and $j\ne k\ne l$ and (5) $\Gamma_5=\Gamma_1\Gamma_2\Gamma_3\Gamma_4$. These 16 matrices all come from the four $\Gamma$ matrices and their products.
The $\Gamma$ matrices satisfy the Clifford algebra of a four-dimensional Euclidean space $\{\Gamma_a,\Gamma_b\}=2 \delta_{ab}$ and
the $\Sigma$ matrices satisfy the so(4) Lee algebra and form a representation for the generators of the SO(4) rotational group.

For systems with time-reversal and space-inversion symmetry ($T^2=+1$), the Hamiltonian can be made real when proper basis are utilized.
For a $4\times 4$ real-symmetric matrix, one can use the following 10 matrices as its basis (1) the identity matrix $I$, (2) $\Gamma_1$, $\Gamma_2$, $\Gamma_3$,
(3) $\Sigma_{14}$, $\Sigma_{24}$ and $\Sigma_{34}$, (4) $i \Gamma_1\Gamma_2\Gamma_4$, $i \Gamma_2\Gamma_3\Gamma_4$, $i \Gamma_1\Gamma_3\Gamma_4$.
With this basis, any Hamiltonian of a four-band system with time-reversal and space-inversion symmetry can be written as the linear combination of these 10 matrices
\begin{align}
H=a \Gamma_1+b_1 \Gamma_2+ b_2\Gamma_3 + c_1 \Sigma_{14}+c_2 i \Gamma_2\Gamma_3\Gamma_4 +\ldots
\label{eq:Hamiltonian}
\end{align}
In this basis, the product of the time-reversal and space inversion symmetry is simply $T I=K$, where $K$ implies complex conjugate.
For reasons that will be discussed below, here we choose to showed the explicit form for 5 of the 10 terms. The coefficients here ($a$, $b_1$, $b_2$, $c_1$, $c_2$ and 5 additional terms that we didn't explicitly show here) are real functions of momentum $\mathbf{k}$. In general, this is a complicated Hamiltonian, which requires 10 independent functions of $\mathbf{k}$ to characterize. Although the eigenvalues and eigenfunctions of this Hamiltonian could be computed in principle, they involve solutions of a quartic eigen-equation, whose analytic formula is extremely complicated in general.

As shown in the Appendix, for an arbitrary four-band Hamiltonian with time-reversal and space-inversion symmetry, we can adiabatically deform the Hamiltonian into one of the following two structures
\begin{align}
H_1=a \Gamma_1+b_1 \Gamma_2+ b_2\Gamma_3,
\label{eq:Hamiltonian1}
\end{align}
and
\begin{align}
H_2=a \Gamma_1+c_1 \Sigma_{14}+c_2 i \Gamma_2\Gamma_3\Gamma_4.
\label{eq:Hamiltonian2}
\end{align}
This adiabatic procedure will not induce  any new level crossing between the two conduction and two valence bands. As a result, the topological structure and topological indices  (e.g. the monopole charge of nodal lines~\cite{Fang:2015aa}) remain invariant. The existence of such an adiabatic deformation provides us an easy pathway to classify possible topological states here. Instead of trying to classify all arbitrary Hamiltonians as shown in Eq.~\eqref{eq:Hamiltonian}, we reduce the problem into classifying the topological structures for the two reduced Hamiltonians $H_1$ and $H_2$ [Eqs.~\eqref{eq:Hamiltonian1} and~\eqref{eq:Hamiltonian2}].

In addition, it is worthwhile to point out that the reduced Hamiltonian $H_1$ and $H_2$ greatly simplifies the topological index calculation because of two reasons. First, in constrast to a generic Hamiltonian, these two reduced Hamiltonians can be easily diagonalized and the eigen-wavefunctions take a simple analytic form, making topological information  easy to access. Secondly, these two Hamiltonians has an extra $Z_2$ symmetry, and as will be shown below, this $Z_2$ symmetry provides a direct connection between
topological indices and high-symmetry-point band inversions, which greatly simplifies the topological index calculation.

As one can easily check,  $H_1$ and $H_2$ share a lot of identical properties. For both of them, the two conduction (valence) bands are degenerate, and the eigen-energy share the same functional form.
\begin{align}
\epsilon_1=\pm \sqrt{a^2+b_1^2+b_2^2}
\end{align}
for $H_1$ and
\begin{align}
\epsilon_2=\pm \sqrt{a^2+c_1^2+c_2^2}
\end{align}
for $H_2$.
These similarities between $H_1$ and $H_2$ are not accidental. Instead, it is the direct consequence of a duality relation, which will be discussed in the following section.

\subsection{Duality relation}
There is a duality relation between $H_1$ and $H_2$, which implies that these two Hamiltonians must share identical algebraic properties.

As mentioned above, the $\Gamma$ matrices form a representation of the Clifford algebra. On the other hand, the exact form of $\Gamma$ matrices is not unique.
For example, we can choose to use another set of $\Gamma$ matrices:
\begin{align}
\tilde{\Gamma}_1&=\Gamma_1
\label{eq:gamma:tilde:1}\\
\tilde{\Gamma}_2&=\Sigma_{14}
\label{eq:gamma:tilde:2}\\
\tilde{\Gamma}_3&=i \Gamma_2\Gamma_3\Gamma_4
\label{eq:gamma:tilde:3}\\
\tilde{\Gamma}_4&=-\Sigma_{12}
\label{eq:gamma:tilde:4}
\end{align}
This new set of $\tilde{\Gamma}$ matrices also follow the same Clifford algebra, and thus are fully equivalent to the original ones that we used above from the algebraic point
of the view.

In other words,  the $\Gamma$ matrices and the $\tilde{\Gamma}$ matrices provide two representations of the Clifford algebra, mathematically
equivalent to each other. More interestingly, these two representations are the dual of each other $\tilde{\tilde{\Gamma}}_i=\Gamma_i$, i.e. the same transformation from
$\Gamma$ to $\tilde{\Gamma}$ [Eqs.~\eqref{eq:gamma:tilde:1}-\eqref{eq:gamma:tilde:4}] also maps $\tilde{\Gamma}$ back to $\Gamma$. To observe this duality relation,
we first use these new $\tilde{\Gamma}$ matrices and their products to define another set of 16 matrices as the basis for $4\times4$ Hermitian matrices, and it is easy to verify that this new set of matrices satisfies the dual relation of Eqs.~\eqref{eq:gamma:tilde:1}-\eqref{eq:gamma:tilde:4}.
\begin{align}
\Gamma_1&=\tilde{\Gamma}_1
\label{eq:gamma:tilde:dual:1}\\
\Gamma_2&=\tilde{\Sigma}_{14}
\label{eq:gamma:tilde:dual:2}\\
\Gamma_3&=i \tilde{\Gamma}_2\tilde{\Gamma}_3\tilde{\Gamma}_4
\label{eq:gamma:tilde:dual:3}\\
\Gamma_4&=-\tilde{\Sigma}_{12}
\label{eq:gamma:tilde:dual:4}
\end{align}
and thus $\tilde{\tilde{\Gamma}}_i=\Gamma_i$.
Furthermore, this duality relation also applies to $H_1$ and $H_2$, i.e. $H_1$ and $H_2$ swaps as we swaps $\Gamma_i$ with $\tilde{\Gamma}_i$.
\begin{align}
H_1=a \Gamma_1+b_1 \Gamma_2+ b_2\Gamma_3=a \tilde{\Gamma}_1+b_1 \tilde{\Sigma}_{14}+b_2 i  \tilde{\Gamma}_2\tilde{\Gamma}_3\tilde{\Gamma}_4,
\end{align}
and
\begin{align}
H_2=a \tilde{\Gamma}_1+c_1 \tilde{\Gamma}_2+ c_2 {\Gamma}_3 = a \Gamma_1+c_1 \Sigma_{14}+c_2 i \Gamma_2\Gamma_3\Gamma_4,
\end{align}

In summary, the duality relation can be represented as
\begin{align}
\Gamma_1& \Leftrightarrow \Gamma_1
\\
\Gamma_2&\Leftrightarrow \Sigma_{14}
\\
\Gamma_3&\Leftrightarrow  \Gamma_2 \Gamma_3\Gamma_4
\\
\Gamma_4&\Leftrightarrow -\Sigma_{12}
\\
H_1&\Leftrightarrow  H_2
\\
P_1&\Leftrightarrow  P_2
\end{align}
where $P_1$ ($P_2$) is a $Z_2$ operator that will be defined in the next section.
Because $\Gamma$ and $\tilde{\Gamma}$ matrices satisfy the same algebra, this duality relation implies that
$H_1$ and $H_2$ must have the same algebraic properties.

\subsection{$Z_2$ symmetry and the parity Chern number}
$H_1$ and $H_2$ also share another important common feature. Both these two Hamiltonians preserve a $Z_2$ symmetry. We can define two $Z_2$ parity operators
\begin{align}
P_1=i \Gamma_1\Gamma_2\Gamma_3
\label{z2p1}
\end{align}
and
\begin{align}
P_2=i \tilde{\Gamma}_1\tilde{\Gamma}_2\tilde{\Gamma}_3
\label{z2p2}
\end{align}
It is easy to verify that
\begin{align}
P_1^2=P_2^2=1
\end{align}
and
\begin{align}
[H_1,P_1]=[H_2,P_2]=0
\end{align}
Thus, $H_1$ ($H_2$) preserves the $Z_2$ symmetry defined by $P_1$ ($P_2$). Notice that $P_1$ and $P_2$ are also dual to each other.

In addition, it must be emphasized that these parity operates anti-commute with $T I$.
\begin{align}
\{P_1,T I\}=\{P_2,T I\}=0
\end{align}
This anti-commutation relation implies two things. First of all, it requires that for $H_1$ or $H_2$, the two valence (conduction) bands must be degenerate. This is because no $1D$ representation can give such an anti-commutation relation, and thus $2$-fold degeneracy has to be required, in analogy to Kramers doublets from the $T^2=-1$ time-reversal symmetry. Secondly, due to the same anti-commutation relation, the two degenerate quantum states here (in either the valence or the conduction bands) must have opposite parity. Here, the parity refers to the +1/-1 eigenvalue of $P_1$ or $P_2$.

This degeneracy/parity property is very important because it allows us to define a topological index, which will be called the parity Chern number. Among the two degenerate valence bands, the parity of this $Z_2$ symmetry allows us to select one band with +1 (or -1) parity eigenvalue. For any 2D closed manifold in the 3D $k$-space (e.g. the $k_z=0$ plane),  we can then compute the Chern number of this band that we selected: $C^+$ for the band with parity $+1$ and $C^-$ for the other band with parity $-1$. This parity Chern number is an integer topological index, in analogy to the spin Chern number or mirror Chern number in quantum spin Hall systems or systems with mirror symmetry.  The time-reversal and space-inversion symmetries here imply that the total Chern number must vanish, and thus $C^+=-C^-$.

For $H_1$ and $H_2$, this parity Chern number has a clear and specific physical meaning. If this topological index is nonzero for any closed and contractible k-space 2D manifold,
then
it implies that this manifold has a nontrivial topological structure and thus cannot be adiabatically shrink into a point. In other words, some gapless nodal points shall
arise in the 3D space enclosed by this manifold, which prevent this manifold from adiabatically shrinking into a point. Generically, these nodal points are Dirac points
with four-fold degeneracy (i.e. the two conduction and two valence band all becomes degenerate at this Dirac point), and each Dirac point carries a parity Chern
number $\pm1$. Thus, the parity Chern number of a 2D manifold here measures directly the number of Dirac points enclosed by this manifold. As will be shown
below, when certain rotational symmetry is enforced, more complicated band crossing points, e.g. quadratic or cubic band crossing points in analogy to their 2D
counterparts~\cite{Sun:2009aa}, can also emerge.  In general, one quadratic (cubic) band crossing point carries a parity Chern number $\pm2$ ($\pm 3$).

As we deviates away from $H_1$ and $H_2$ (but preserves the time-reversal and space-inversion symmetry), the $Z_2$ symmetry defined by $P_1$ or $P_2$ is explicitly broken, but these Dirac points cannot be gapped out. Instead, they develop into topological nodal lines. In particular, these topological nodal lines are not conventional ones. Instead, each of them carries a nontrivial monopole charge.  This is because $C^+ \textrm{ mod } 2$ coincides with another topological index, the $Z_2$ monopole charge introduced in Ref.[\onlinecite{Fang:2015aa}]. As we deviate the Hamiltonian from $H_1$ and $H_2$,  the $Z_2$ symmetry will in general be explicitly broken and thus the parity Chern number is no long well defined,
but the parity of the parity Chern number (i.e. the monopole charge) remains as a good topological index. For any adiabatic deformation of the Hamiltonian, this index remains invariant and thus a Dirac point for $H_1$ and $H_2$ with $C^+=\pm 1$ shall evolve into a topological nodal line with monopole charge $1$.

In general, as shown in Ref.[\onlinecite{Fang:2015aa}], a monopole charge is a $Z_2$ index with value being either $0$ or $1$. However, as will be shown below, if additional symmetry is enforced, monopole charge may take higher values beyond $1$. The physical meaning of a higher monopole charge is two fold. (1) In certain cases, it is associated with the number of topological nodal lines. As we mentioned early on, for $H_1$ and $H_2$, the value of parity Chern number provides the number of Dirac points. In the presence of certain rotational symmetry, we may have multiple Dirac points (associated with each other via some symmetry transformation), and thus a large-than-one parity Chern number. As we deviate from $H_1$ or $H_2$, each of the Dirac points will develops into a topological nodal line, and thus the parity Chern number provides us direct information about the number of topological nodal lines that we should expect in the Brillouin zone. (2) In certain other cases, as will be shown below, a higher index doesn't produce multiple nodal lines. Instead, we found one single topological nodal line carriers high topological index (monopole charge 2) in systems with four-fold or six-fold rotational symmetries. These nodal lines must be around a high symmetry axis and preserve the corresponding rotational symmetry, and their high index can be justified by defining a $Z_2$ monopole charge in a fraction of a closed 2D manifold embedded
in the 3D Brillouin zone.

Another key property of these parity Chern number lies in the fact that it can be directly associated with band inversions at high symmetry points, through a technique derived in Ref.[\onlinecite{cfang2012}]. Thus, our topological index offers a bridge way to connect band inversions at high symmetry points with the topological nature of nodal lines. With this bridge, in most cases, information about high symmetry points can fully dictate topological nodal lines, especially those with nontrivial monopole charge. This result is one key observation of this manuscript, and it allows us to classify and locate topological nodal lines (with nontrivial monopole charge) purely based on local information of the band structure.

\subsection{Theoretical approach}
Below, we will utilize the techniques discussed above to analyze and classify topological nodal lines in 3D systems with time-reversal and space-inversion symmetries. We will consider systems with different rotational symmetries and various types of band inversions. As mentioned above, here, we first focus on four-band models with two conduction/valence bands, but  the topological indices discussed here is well defined in generic systems with arbitrary number of bands and thus all the topological properties remain invariant, when extra bands are introduced. In addition, for simplicity, we will focus on the case where band inversions only occur at the $\Gamma$ point ($\mathbf{k}=0$), but the same discussion and principle can be generalized to deal with band inversions at other high symmetry points.

We will classify the band inversion based on symmetry and the point group representation of each band. For each allowed band inversion type, we will compute the parity Chern number (defined above) for $H_1$  and $H_2$ over the $k_z=0$ plane and the $k_z=\pi$ planes. Because we have shown above that any four-band Hamiltonian can be adiabatically deformed into either $H_1$ and $H_2$, although we only examine $H_1$ and $H_2$ in this effort, our conclusions generically apply to any generic Hamiltonians. With this parity Chern numbers, we can then define a topological index
\begin{align}
Q=C^+_{k_z=\pi}-C^+_{k_z=0}
\label{eq:topological_index_Q}
\end{align}
This index measures the difference between the parity Chern number in the $k_z=0$ and $k_z=\pi$ planes. For $H_1$ and $H_2$, this index will be labeled as $Q_1$ and $Q_2$ respectively.

As mentioned in the previous section, for the Hamiltonian $H_1$ or $H_2$, this index $Q$ measures the number of Dirac points located between the  $k_z=0$ and $k_z=\pi$ planes, i.e. the upper half of the 3D Brillouin zone ($0<k_z<\pi$). It is easy to verify that the lower half of the Brillouin zone ($-\pi<k_z<0$) have the same number of Dirac points due to the
space-inversion symmetry. For a generic Hamiltonian (away from $H_1$ or $H_2$), each Dirac points will evolve into a topological nodal line (with a nontrivial monopole charge), and thus this index $Q$ measures the number of topological nodal loops (with nontrivial monopole charge) in the upper half of the Brillouin zone.  This is how we determine the topological structure and classify topological states in these systems. In particular, for the simplest case (without any other symmetries), $Q \textrm{ mod } 2$ is just the monopole charge of the upper half Brillouin zone.

To compute this topological index  $Q$, we applies the method developed in Ref.[\onlinecite{cfang2012}] for the $k_z=0$ and $k_z=\pi$ planes respectively. As shown below and in the Appendix, this technique allows us to determine $Q$ purely using information about high-symmetry-point band inversions, up to modular $2n$ for a system with $2n$-fold rotational symmetry or $4n+2$ for system with $(2n+1)$-fold rotational symmetry. These values are fully dictated by what happened at high symmetry points, while microscopic details
away from high symmetry points become irrelevant.

\begin{figure}
\includegraphics[width=0.3\textwidth]{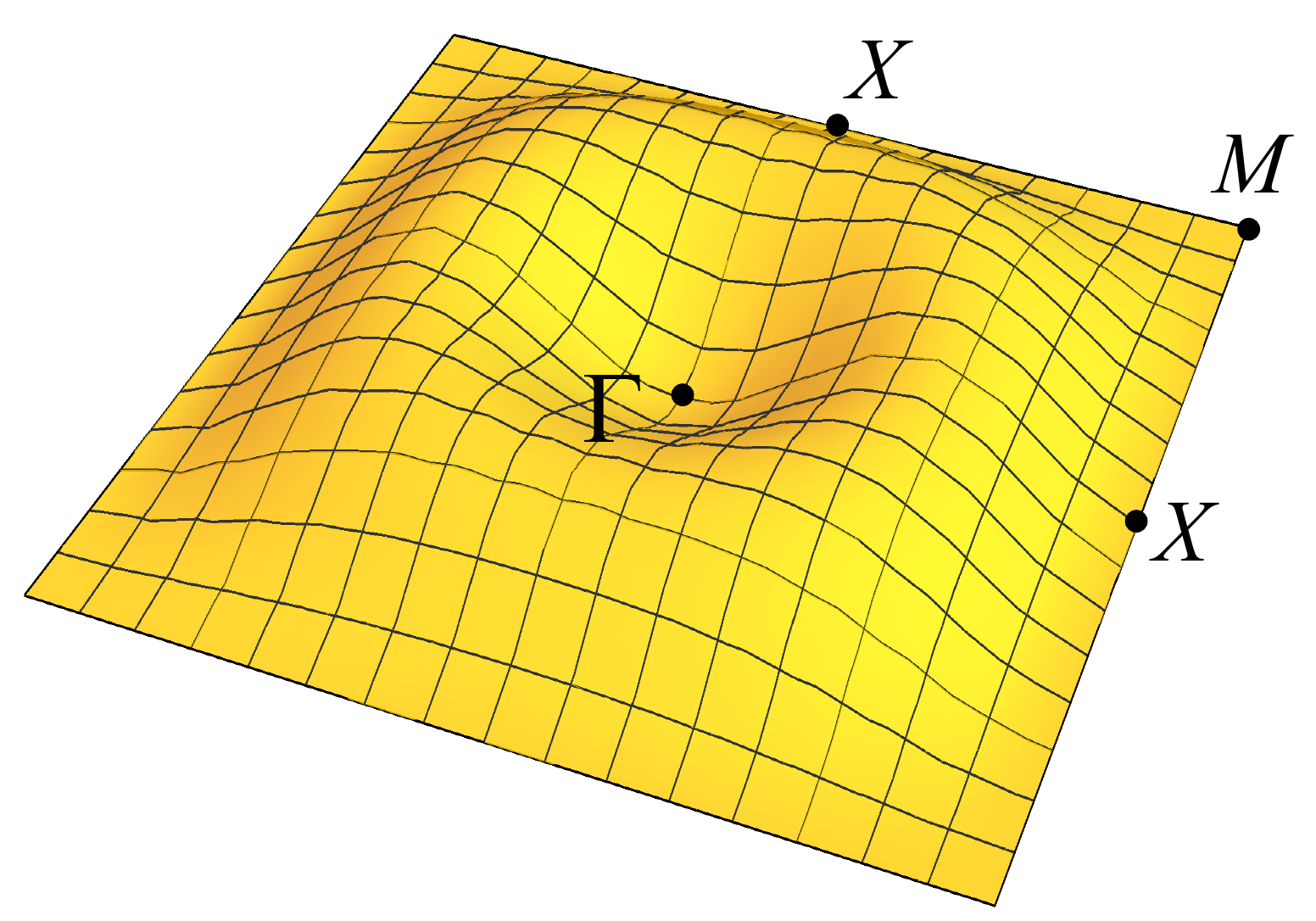}
\caption{A deformed $k_z=0$ plane for the calculation of topological indices. If the $k_z=0$ plane is not fully gapped, we can choose a slightly deformed plane to calculate the topological index. Here we demonstrate one example by considering a system with four-fold rotational symmetry around the $z$ axis. With inversion symmetry and four-fold rotational along $z$, the $k_z=0$ plane is a mirror plane ($k_z \rightarrow -k_z$), and thus it may contain mirror-symmetry-protected topological nodal lines. To find a fully gapped 2D manifold in the $k$-space to define the topological index (e.g. $C^+$ or the monopole charge), we can deform a bit from the $k_z=0$ plane as shown in the figure. The deformed 2D manifold still contains all the four high symmetry points of the original $k_z=0$ plane ($\Gamma$, $X$ and $M$), and it still preserves the four-fold rotational symmetry along $z$. This 2D manifold allows us to avoid nodal lines or nodal points in the $k_z=0$ plane and thus topological indices can be defined and computed using high symmetry points.}
\label{fig:curved}
\end{figure}

As will be shown below, in certain cases, the $k_z=0$ or $k_z=\pi$ planes are not fully gapped, e.g. mirror symmetry-protected nodal lines may arise if $k_z=0$ or $\pi$ is a mirror plane. For these cases, we will choose a 2D plane in the Brillouin zone slightly deformed from $k_z=0$ or $\pi$, as demonstrated in Fig.~\ref{fig:curved}. This deformed plane respects the same rotation symmetry (for rotations along the z axis), and it contains all the four high symmetry points of the $k_z=0$ or $\pi$ plane, i.e. $k_x$ and $k_y$ being $0$ or $\pi$. This deformed k-space 2D plane will avoid the gapless points or nodal lines, such that we can define and compute the parity Chern number mentioned above. Because we preserved the high symmetry points and the rotational symmetry, the parity Chern number can still be dictated by high-symmetry-point band inversions same as before.

We should also emphasize that here we only focus on generic topological properties that are robust against fine tuning or perturbations, as long as the symmetry and the gap remain. Thus, in general, it is possible that additional nodal lines or nodal points may arise, e.g. accidental degeneracy, but those features in general can be gapped out via adiabatic perturbations and would not be considered in our study.

\subsection{Comparison with lattice models}
In addition to theoretically analyzing topological indices using reduced Hamiltonian $H_1$ and $H_2$, we also constructed lattice models with generic Hamiltonians beyond $H_1$ and $H_2$. These lattice models are constructed based on symmetry without further constraints. For a specific lattice (e.g. tetragonal), we introduces orbits with different symmetry to each Bravais lattice site, and systematically included all symmetry-allowed short-range hopping terms in our model Hamiltonian (Details about the symmetry-based construction can be found in Ref.[\onlinecite{Dresselhaus:2010aa}]). In these lattice models, we randomly assign values to the control parameters (e.g., the hopping strength) and compute the band structure by numerical diagonalizing the Hamiltonian in the k-space. Then we compared nodal lines and other topological features that are observed in these models with our generic theoretical classifications (using $H_1$ and $H_2$). The generic classification and model calculations agree with each other in all the cases that we have examined.

\begin{figure*}[t]
\includegraphics[width=6.5in]{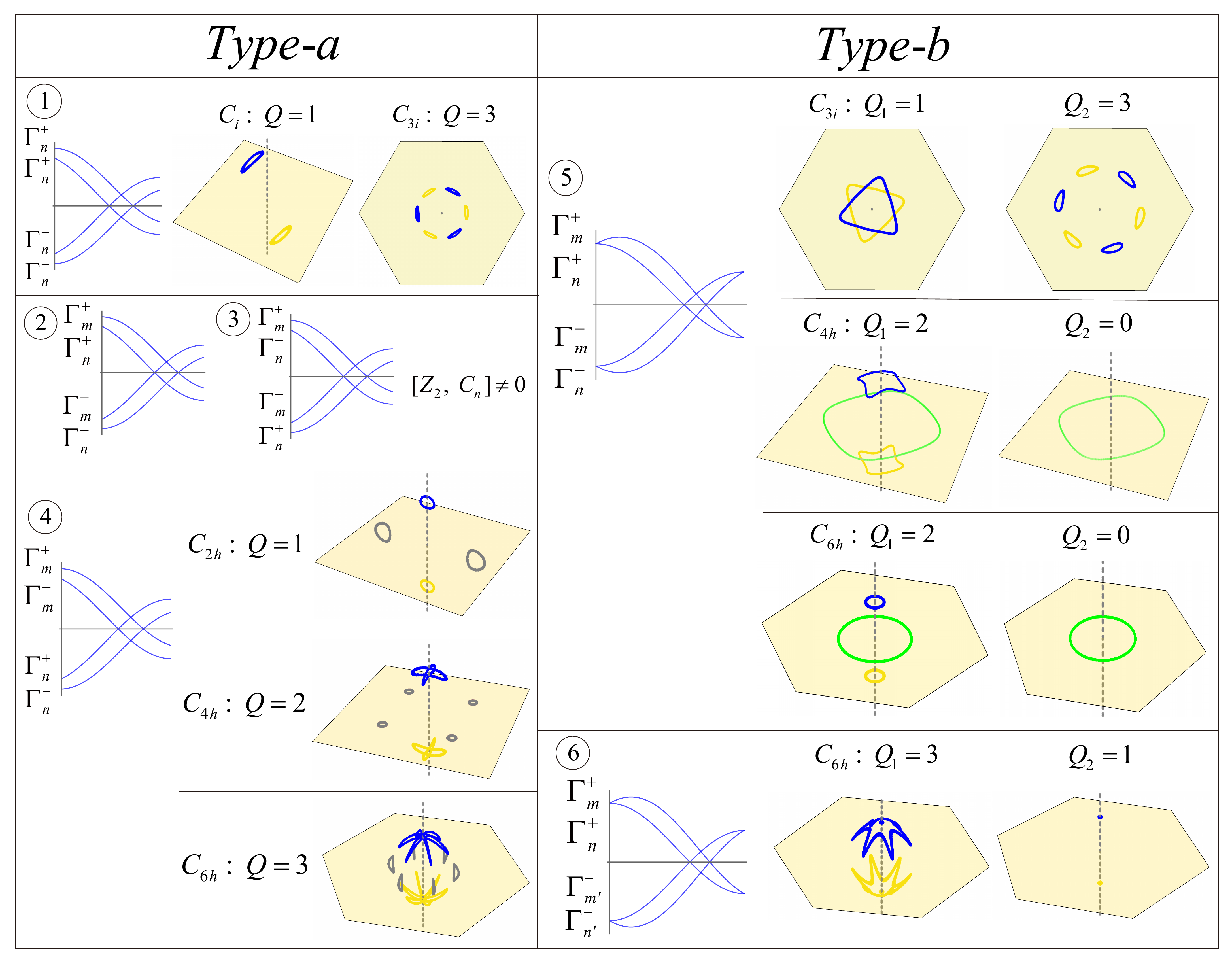}
\caption{Band inversions and topological nodal lines in systems with various symmetries. Blue (yellow) nodal lines represent nodal lines in the upper (lower) half of the Brillouin zone with $0<k_z<\pi$ ($-\pi<k_z<0$). All these nodal lines carrier nontrivial monopole charge and thus they cannot be gapped out by shrinking into a point. Green nodal lines are protected by the mirror symmetry and is located in the $k_z=0$ plane. Gray nodal lines cannot be detached from the $k_z=0$ mirror plane, but they are not mirror-symmetry-protected nodal lines. Instead, these gray nodal lines also carry nontrivial monopole charge and thus cannot be gapped out by shrinking into a point.}
\label{table}
\end{figure*}

\section{Band inversions in the presence of time-reversal symmetry}
In this section, demonstrate the generic techniques discussed above by focusing on systems with one pair of bands inverted at the $\Gamma$ point, and no band inversion
takes place at any other high symmetry points. We will first provide a summary of result, while details are being provided afterwards.

\subsection{Summary of results}
As mentioned early on, we consider systems with time-reversal
and space-inversion symmetry, and ignore spin-orbital effects ($T^2=+1$). In addition, we assume that the system preserves the $n$-fold rotational symmetry along the $z$ axis with $n=1, 2, 3, 4, 6$. We focus on systems with a double band inversion at the $\Gamma$ point, which can be classified and labeled according to the symmetry representations of
the bands as summarized in the table shown in Fig.~\ref{table}. Here we distinguish two different scenarios  depending on whether the inverted bands at $\Gamma$ belongs
to type-a or type-b representations, and 6 different families of band inversions are examined here as shown in the table. Here, we follow the same notation
utilized in the study of point or space groups~\cite{pointgroup}, where type-a and type-b represent two different types
of response under time-reversal (See below for their definitions and other details).

For band inversions between type-a bands, it turns out that high-symmetry-point band inversions fully dictate the topological index and the configuration of the topological nodal lines, while for type-b bands, for each specific band inversion, two quantum phases with distinct topological structure and nodal line configurations are always allowed. These two phases are separated by a topological transition,  signatured by a gap closing in (or near) the $k_z=0$ plane.

In particular, we would like to highlight band inversion family 4, 5 and 6 shown in Fig.~\ref{table}. For family 4, we found a special type of nodal lines (blue and yellow lines) which are oriented along a high symmetry axis. Each of these nodal lines carry the monopole charge $1$. Remarkably, for systems with a four-fold or  six-fold rotational symmetry, these nodal lines have to tangle together to form a cage like structure, due to a combined constraint from topology and symmetry. Because these nodal lines are connected, the monopole
charge of each line becomes ill defined. Instead, the correct topological index here is the monopole charge of a cage structure, which carries a monopole charge $2$
and $3$, if the rotational symmetry is four- and six-fold respectively. In addition, this family may also show nodal lines near the $k_z=0$ plane (gray lines).
These nodal lines may orient parallel or perpendicular to the $k_z=0$ plane and they cannot be detached from the $k_z=0$ plane due to the mirror symmetry ($k_z\rightarrow -k_z$). Although pinned by the mirror symmetry, these nodal lines are not the conventional mirror-symmetry protected nodal lines. Instead, each of them carrier
a nontrivial monopole charge $1$.

In family 5 and 6, for systems with a four- or six-fold rotational symmetry, we observe another new type of nodal lines (blue and yellow). Each of these nodal lines carries a monopole charge $2$, in direct contrast to nodal lines without rotational symmetry, whose the monopole charge is a $Z_2$ index and thus can only be $0$ or $1$. These monopole-charge-2 nodal lines must be oriented around a high symmetry axis (the $k_z$-axis). As will be shown below, the rotational symmetry enables us to define the monopole charge on half of a 2D closed manifold in these cases, and the rotational symmetry requires both of the half manifolds share the same monopole charge. As a result,
for each half manifold, the monopole charge is still a $Z_2$ topological index ($0$ or $1$), but for the whole manifold, the monopole charge shall take the value $0$ or $2$ after adding up the contribution from both half pieces. For a monopole charge $1$ nodal line, we know that it can be adiabatically  shrunk in to a Dirac point with linear dispersion. For these monopole charge $2$ nodal lines, they will shrink into quadratic (or even cubic) band crossing points with quadratic dispersion in directions perpendicular to the rotational axis and linear  along $k_z$.

\subsection{Three types of time-reversal symmetry in lattice systems}
For a homogenous system with $SO(3)$ rotational symmetry, time-reversal symmetry can be classified into two categories, depending on whether we have integer or half-integer spins. In a lattice systems, however, this classification becomes more complicated due to the interplay between space and time symmetry. Instead of two categories, three categories becomes needed, dubbed type-a, b and c.~\cite{pointgroup}

A type-a representation of a point group is also known as a real representation. Under time-reversal, a type-a representation is mapped back to itself. Type-b representations are also known as complex representations. Under time-reversal, a type-b representation will turn into another representation, different from the original one. For example, in a system with
three-fold rotational symmetry (point group $C_3$), this group has three presentations doubled $\Gamma_1$, $\Gamma_2$ and $\Gamma_3$. Under a  $C_3$ rotation, states belongs to the $\Gamma_1$ representation remains unchanged, while states in $\Gamma_2$ or $\Gamma_3$ representation pick up a complex phase $e^{2\pi/3 i}$ or $e^{-2\pi/3 i}$ respectively. Here, $\Gamma_1$ is a type-a representation, who is its own time-reversal partner, while $\Gamma_2$ and  $\Gamma_3$ are type-b representations and they are the time-reversal partner of each other. This can be easily checked by realizing that due to its anti-Hermitian nature, time-reversal flips the complex phase factors $e^{2\pi/3 i}$ and $e^{-2\pi/3 i}$, which is why $\Gamma_2$ and $\Gamma_3$ are connected by a time-reversal transformation. Type-c representations are similar to type-a in the sense that the representation is its own time-reversal partner. However, in contrast to type-a, type-c representations cannot be made real due to the half-integer spin quantum number. In this paper, we will only consider type-a and type-b representations, because type-c representations are prohibited in systems without spin-orbit couplings ($T^2=+1$).\cite{pointgroup}

In comparison to a homogenous system with $SO(3)$ rotational symmetry, type-a and type-c corresponds to integer and half-integer spins respectively.  In contrast, type-b is the most special one among the three, and it has no correspondence in the $SO(3)$ rotational group.  Remarkably, we found that band inversions between between type-a bands and band inversions between type-b bands result in totally different topological structures and topological classifications, which is one of the key results of this manuscript as
summarized in the Table in Fig.~\ref{table}.


\subsection{Point group $C_i$}
\begin{figure}
\subfigure[]{\includegraphics[width=0.2\textwidth]{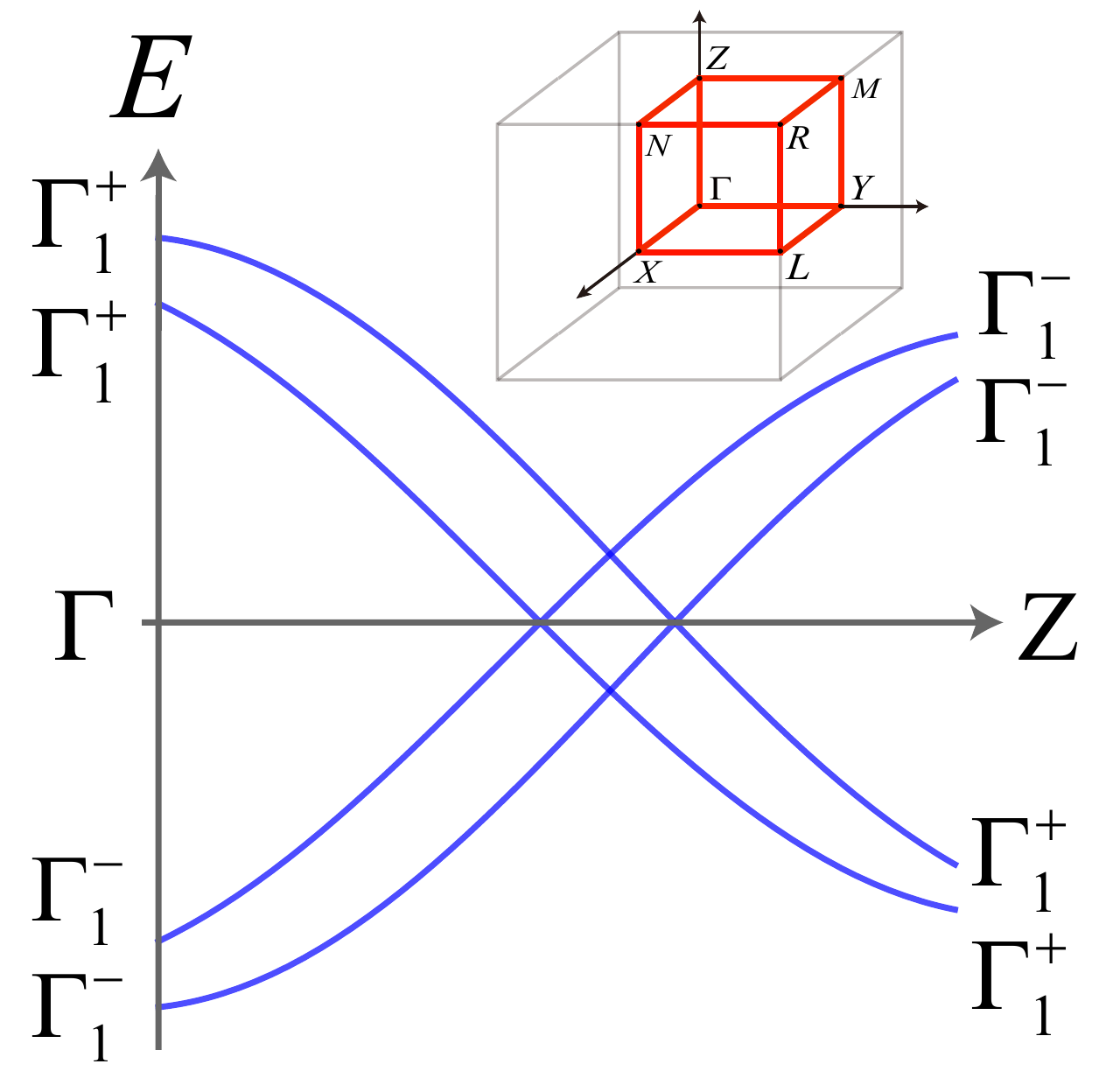}}
\subfigure[]{\includegraphics[width=0.25\textwidth]{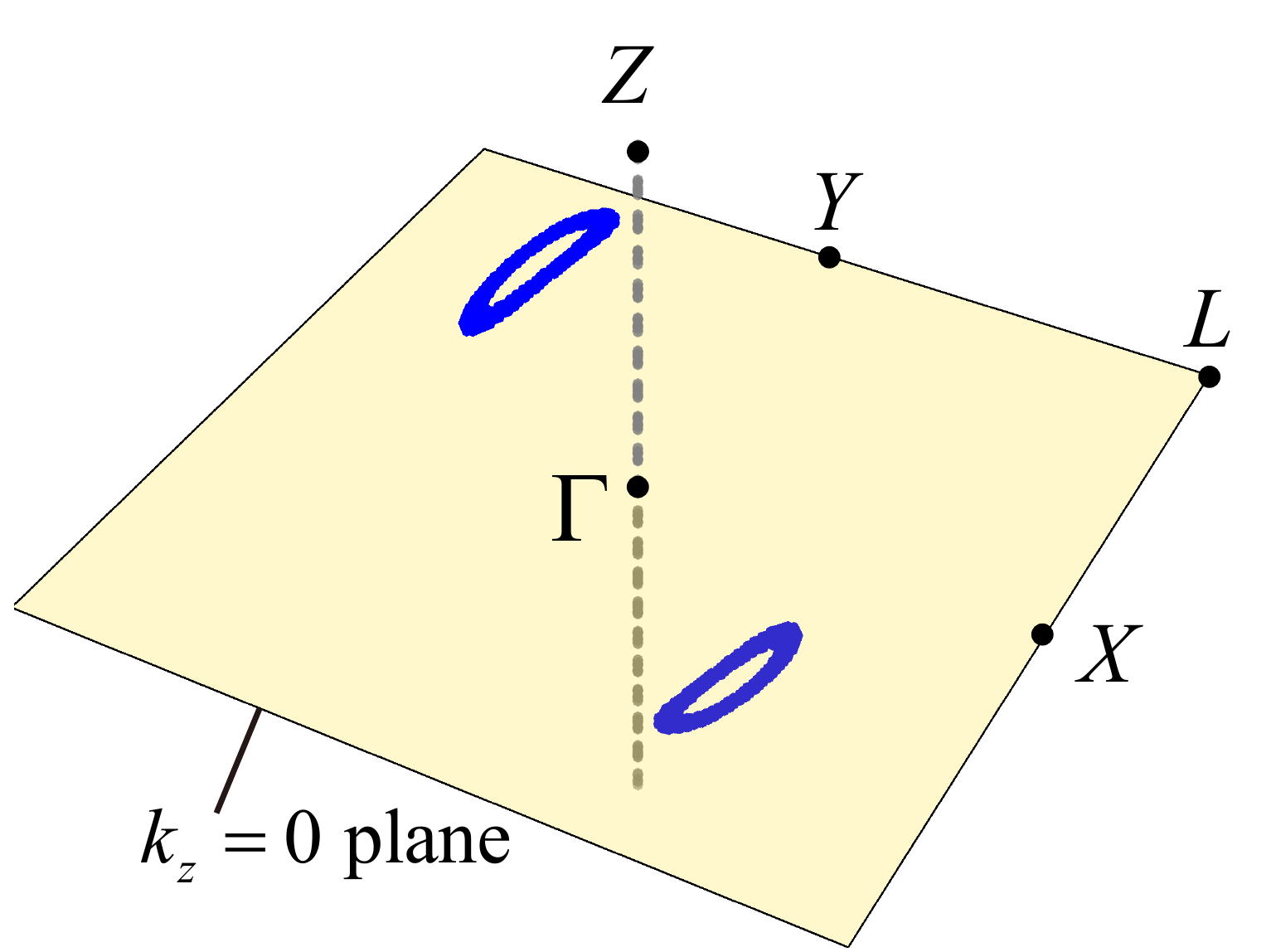}}
\caption{(a) Schematic band structure demonstrating a pair of band inversion at $\Gamma$. The inset shows the Brillouin zone of a system with $C_i$ symmetry. Here, the eight high symmetry points are labeled as $\Gamma$, $X$, $Y$, $L$, $Z$, $M$, $N$, $R$ following the convention of a triclinic Bravais lattice.  At $\Gamma$, the two valence (conduction) bands have the $-$ ($+$) parity under space inversion, which is opposite to all other high symmetry points where the valence (conduction) bands have the $+$ ($-$) parity.
(b) Topological nodal lines induced by this band inversion.}
\label{fig:Ci_inversion}
\end{figure}

If the system has no other symmetry beyond space-inversion and time-reversal, the crystal point group is $C_i$. Without spin-orbit coupling, the group $C_i$ has two representations, $\Gamma_1^+$ and $\Gamma_1^-$ (even or odd under space inversion respectively), and both these two representations are type-a. For this case, all the eight high symmetry points share the same point-group symmetry.  With the assumptions presented above, only one type of band inversion is
possible here as shown in Fig.~\ref{fig:Ci_inversion}. At the $\Gamma$ point, the two valence bands belong to the $\Gamma_1^-$
representation and the two conduction bands has the symmetry of $\Gamma_1^+$, while all other high symmetry points are opposite
with the valence (conduction) bands belonging to the $\Gamma_1^+$ ($\Gamma_1^-$) representation.

Here, we can look at the $k_z=0$ plane. For $H_1$, we use the $Z_2$ symmetry $P_1$ to define a parity Chern number.
Because for the 2D $k_z=0$ plane, space inversion is effectively equivalent to a $\pi$ rotation along the $z$-axis, this $k_z=0$ plane can be considered as a 2D system with 2-fold rotational symmetry along $z$. As shown in Ref.[\onlinecite{cfang2012}], with 2-fold rotational symmetry, we can compute a Chern number up to mod 2 using high symmetry points rotational eigenvalues (i.e. $+1/-1$ parity of the valance bands for our system here). Here, because the rotational operator commute with the $Z_2$ symmetry, we can apply the same technique for the parity Chern number and find that $C^+_{k_z=0}\mod 2=1$. The same analysis for the $k_z=\pi$ plane produces $C^+_{k_z=\pi} \mod 2=0$.
Thus, $Q_1\mod 2=1$, where $Q$ is the topological index defined above in Eq.~\eqref{eq:topological_index_Q} and the subindex $1$ here implies that it is for $H_1$. For $H_2$, we can perform the same analysis and we find that $Q_1=Q_2=1$ up to mod $2$.

Because any Hamiltonian is adiabatically connected with either $H_1$ or $H_2$, the fact $Q_1=Q_2=1$ implies that this band inversion always produces the same topological structure, with $Z_2$ monopole charge $1$. For $H_1$ and $H_2$, as mentioned above, the topological index implies that one Dirac point is located in the upper half of the Brillouin zone ($0<k_z<\pi$), and one in the lower half ($-\pi<k_z<0$). For a generic Hamiltonian deviated from $H_1$ or $H_2$, each Dirac point evolves into a topological nodal line with a nontrivial $Z_2$ monopole charge, and thus we expected one nodal line located in the upper half of the Brillouin zone  and one in the lower half. Both these two nodal lines carry nontrivial monopole charge, and thus they cannot be gapped out by shrinking a nodal loop into a point.~\cite{Fang:2015aa}

As shown in Fig.~\ref{fig:Ci_inversion}, we setup a four-band lattice model on a triclinic lattice with four orbits (two with $+$ and two with $-$ parity under space inversion) at each Bravais
lattice site. With this band inversion, the band structure indeed shows nodal lines and monopole charge as predicted above. Details about the model can be found in the Appendix.

It is worthwhile to mention that the index $Q$ here is determined up to mod 2, and thus for $H_1$ and $H_2$ the number of Dirac points is determined only up to mod 2. In principle, extra pairs of Dirac points may arises in the upper half of the Brillouin zone (and the same number of pairs of Dirac points will arise in the lower half as well due to the inversion symmetry). These extra pairs are allowed but not generic or universal, because in general, a pair of Dirac points can annihilate with each other and be gapped out via adiabatic deformations. Thus, because we have confined our to focus on generic and universal properties as stated in earlier sections, these extra pairs of nodal lines will not be consider, neither will we try to obtain the value of $Q$ beyond mod 2 in this symmetry class.

\subsection{Systems with $n$-fold rotational symmetries}
If the system preserves a $n$-fold rotational symmetry along $z$ ($n>1$), in addition to space-inversion and time-reversal symmetries, more complicated band inversions become allowed. Here, again we focus on cases with one pair of bands inverted at the $\Gamma$ point. And as highlighted in the Introduction, we consider small inversion
at $\Gamma$, where the inverted gap is small compared with other energy scales in the system.
The results is shown in Fig.~\ref{table}, where we showed a table summarizing 6 families of possible band inversions. It is worthwhile to emphasize that here we didn't try to exhaust all possible types of band inversions, but the same technique can be easily generalized.

One remarkable observation here lies in the fact that if the inverted bands have type-a time-reversal symmetry, the band inversion fully dictates the value of the topological index. However, if the inverted bands all have type-b time-reversal symmetry, then the topological indices may take two different values, i.e. the band inversion doesn't fully dictate the topological index. In these systems, as we tune the values of control parameters, a topological phase transition will be triggered, at which some of the nodal lines annihilate and reconfigure with each other.

Same as before, the band inversion is characterized by the symmetry properties of the inverted bands, and labeled by the symmetry representations of each band at high symmetry points. Here, we follow the group theory notation used in Ref.[\onlinecite{pointgroup}]. In general, for systems with the space-inversion symmetry, a representation of a point group is labeled as
$\Gamma$ with two indices $\Gamma_n^{\pm}$.
The subindex $n$ is the an integer. Representations with different subindex $n$ show different response to space rotations and mirror reflections, etc.
The superscript $+$ or $-$ labels the
parity under space inversion ($\pm 1$).

\subsection{Band inversion between type-$a$ bands}
In this section, we examine band inversion between bands with  type-$a$  time reversal symmetry.

Here, we start from family 1, where we have four orbits per unit cell, and their symmetry representations all share the same subindex $\Gamma_n$ but two of them have $+$ parity under space inversion, while the other two are $-$. Because these four bands have two distinct parity eigenvalues, we can distinguish them and thus define a pair of band inversion at $\Gamma$. As shown in the
schematic band structure in Fig.~\ref{table}, for family 1, we assume that at the $\Gamma$ point, the two bands with $-$ parity has lower energy, while at all other high symmetry points, it is the two $+$ parity bands who have lower energy. It is worthwhile to mention that in general, $\Gamma$ point preserves the same point group symmetry as the crystal itself, and thus, the bands can be labeled using the representation of the point group of the crystal. However, for other high symmetry points, their point group (known as the little group at that momentum point) is usually  a subgroup of the point group of the crystal. As a results, we should use the representation of the corresponding subgroup to label the bands at those momentum points. For a small band inversion considered here,
the representation for bands away from the $\Gamma$ point can be deduced easily from the point-group compatibility table~\cite{pointgroup}. We will not dive into the mathematical details here, but nevertheless, for any band inversion, the point group allows us to immediately determine the symmetry representations at any high symmetry points. And thus, we can use the formula developed in Ref.[\onlinecite{cfang2012}] to compute the topological index. For band inversions in family 1, it turns out that the index is always trivial, if the system have even-fold rotational symmetry along $z$ ($C_{2n}$). For odd-fold ($C_{2n+1}$), the index is $Q=2n+1$ regardless of microscopic details.  This result implies that with even-fold rotational symmetry, this band inversion will not leads to any nodal lines with nontrivial monopole charge. Instead, it is quite straightforward to note that because even-fold rotation and space inversion imply that the $k_z=0$ plane is a mirror plane, this band inversion will result in a mirror-symmetry-protected nodal line in the $k_z=0$ plane. For systems with odd-fold rotational symmetry ($C_{2n+1}$), $k_z=0$ is not a mirror plane and thus no mirror-symmetry-protected nodal line will emerge, but the nontrivial index $Q$ implies that we shall see $2n+1$ nodal lines in the upper half of the Brillouin zone ($0<k_z<+\pi$), each of which carries a nontrivial monopole charge $1$, and they are related to each other by the $C_{2n+1}$ rotation along $z$. The inversion symmetry implies that another set of $2n+1$ nodal lines shall arise in the lower half of the Brillouin zone $-\pi<k_z<0$.

In family 2 and 3, the bands involved in the band inversion has different rotational symmetry (i.e. their representations have different subindex label $n \ne m$). In this case, high symmetry point doesn't provide direct information about the topological index. This is because for this case, the $Z_2$ symmetry ($P_1$ or $P_2$) dosen't commute with the rotational symmetry ($C_n$). As a result, we cannot apply the method in Ref.[\onlinecite{cfang2012}] to determine the parity Chern number using high-symmetry-point rotational eigenvalues.

In family 4, we consider a different type of band inversion with $n\ne m$. It turns out that this case only arises for systems with even-fold rotational symmetry (2, 4 or 6), where systems with $C_{3i}$ symmetry (three-fold) don't have two different a-type representations and thus we cannot get $n\ne m$. The topological index $Q$ for this case is $n$, if the system preserves $2n$-fold rotation symmetry along $z$. As mentioned above, generically, $Q=n$ implies that for $H_1$ or $H_2$, we will have $n$ Dirac points in the upper (or lower) half of the Brillouin zone. For a generic Hamiltonian (deviated from $H_1$ or $H_2$), each Dirac point evolves into a topological nodal line, and thus $n$ nodal lines should be expected for the upper (or lower) half of the Brillouin zone.

As will be shown below, this naive picture is indeed correct, but the $2n$-fold rotational symmetry here gives us an interesting twist. In fact, the topological index $Q=n$ and the $2n$-fold rotational symmetry together tells us that each topological nodal line here must preserve the $C_2$ symmetry (two-fold rotational along $z$), and thus these nodal lines must organize themself into a cage-like structure as shown in Fig.~\ref{table}.

For $H_1$ and $H_2$, if a system has a $2n$-fold rotational symmetry and we have a Dirac point at momentum $\mathbf{k}$, Dirac points shall also be found at momentum $C_{2n}\mathbf{k}$, $C_{2n}^2\mathbf{k}$ $\ldots$ $C_{2n}^{2n-1}\mathbf{k}$ due to the rotational symmetry. Unless $\mathbf{k}$ is on certain high symmetry lines (e.g. the $k_z$ axis with $k_x=k_y=0$ ), these $2n$ momentum points are in general different from each other and thus $2n$ Dirac points shall be expected in the upper half of the Brillouin zone. However, the topological index dictates that only $n$ Dirac points can exists here, which is half of $2n$. This implies immediately that our Dirac points must be located on certain high symmetry lines, such that the following $2n$ momentum points are not all distinct: $\mathbf{k}$, $C_{2n}\mathbf{k}$, $C_{2n}^2\mathbf{k}$ $\ldots$ $C_{2n}^{2n-1}\mathbf{k}$. Similarly, for the nodal loops, due to the same reason, we should generically expect $2n$ of them in the upper half of the Brillouin zone. However, the fact that $Q=n$ implies that we can have only $n$ nodal lines, and thus each of our nodal line must be invariant under two-fold rotational along $z$ (i.e. under such a rotation, a nodal is mapped back to itself). These combined constraints from topology and symmetry require that the nodal lines must form the cage like structure as shown in Fig.~\ref{table}.

This theory prediction is indeed what we observed in lattice models.  In this case, without utilizing any topological knowledge, the band inversion and the n-fold rotational symmetry along the z-axis automatically implies that two gapless nodal points must arise along the $k_x=k_y=0$ axis. This is because the band inversion cannot satisfy the symmetry compatibility relation of the point/space group along the $z$ axis~\cite{pointgroup}. In other words, along the $k_z$ axis ($k_x=k_y=0$), bands from representations $\Gamma_m$ and $\Gamma_n$ have different symmetry under z-axis rotation, and thus hybridization between them is prohibited. As a result, this band inversion at $\Gamma$ must lead to symmetry protected band crossings along this high symmetry axis. However, it turns out that this symmetry-based conclusion is not the full story. Instead of two individual nodal points, what really arise here are actually cage-like topological nodal loops as shown in  Fig.~\ref{table}. The number of the loops is determined by the rotational symmetry (n loops in the upper half Brillouin zone if we have 2n-fold rotational symmetry). These nodal loops cross with the $k_x=k_y=0$ axis at two points, which are the two gapless points predicted by the symmetry arguments above. And it is easy to verify that each of the topological nodal line here is invariant under a $C_2$ rotation along $z$. As we deform the Hamiltonian towards $H_1$ or $H_2$, these cage-like nodal lines all shrink into a point (on the $k_z$ axis). For system with $2$-fold rotation, this nodal point is a Dirac point, but if we have $4$-fold or $6$-fold rotational symmetry, we will find a quadratic or cubic band crossing point instead. Here, for a quadratic or cubic band crossing point, the dispersion is linear along $k_z$, but in the $k_x$ or $k_y$ direction, the dispersion is quadratic or cubic. These higher-order band crossing point is a 3D version of the 2D quadratic or cubic band crossing point. Similar to their 2D counterparts, a 3D quadratic or cubic band crossing point require rotational symmetries (4 or 6 fold)~\cite{Sun:2009aa}. Otherwise they will split into $2$ or $3$ Dirac points. For the topological index, a quadratic (cubic) band crossing point carriers parity Chern number 2 (3), which is twice (triple) the value of a Dirac point, which is also the reason why it develops into $2$ or $3$ topological nodal lines as we deviates from $H_1$ or $H_2$.

In addition, in family 4, some additional topological nodal lines can arise near the $k_z=0$ plane (gray lines shown in the figure). These nodal lines preserve the mirror symmetry ($k_z \rightarrow -k_z$), but they are  not the conventional mirror-symmetry-protected nodal lines. As we can see from the figure, these nodal lines can extend out of the  ($k_z=0$) mirror plane, as long as it remains invariant under mirror reflection. Each of these nodal lines carries a nontrivial monopole charge $1$.

\subsection{Band inversion between type-$b$ bands}
In family 5 and 6, we consider band inversions from type-b bands, which can occur in systems with 3-, 4- or 6-fold rotational symmetry (i.e. point group $C_{3i}$, $C_{4h}$, $C_{6h}$). At high symmetry points, a band with type-b symmetry must be two-fold degenerate (the band and its time-reversal partner). As a result, a band inversion of type-b bands must invert two bands simultaneously.

For band inversions of type-$b$ bands, it turns out that the nodal line structure in this case is {\it not} fully dictated by band inversions at high symmetry points. For each specific band inversion,
there are always two possible quantum phases with different topological structures. More precisely, we can show that the topological index $Q$ defined above always takes different values for Hamiltonians $H_1$ and $H_2$ ($Q_1\ne Q_2$). Because we have proved above that any Hamiltonian is adiabatically connected to one of these two Hamiltonians ($H_1$ and $H_2$), the fact that they have different topological indices implies that for systems with band inversions involving type-$b$ bands, there are always two different quantum phases. Any quantum systems with Hamiltonian adiabatically connected with $H_1$ belongs to one quantum phase, and all other systems with Hamiltonian smoothly connected with $H_2$ belongs to the other quantum phase. These two quantum phases are separated by a quantum phase transition, at which the number of topological nodal lines will change.

In family 5, the valance bands  at $\Gamma$ involves one pair of time-reversal partners with symmetry representation $\Gamma_m^-$ and $\Gamma_n^-$ ($T \Gamma_m \rightarrow \Gamma_n$ and $T \Gamma_n \rightarrow \Gamma_m$).  For the conduction bands, the corresponding symmetry representation have the same subindices but the parity (under space inversion) is opposite $\Gamma_m^+$ and $\Gamma_n^+$. At all other high symmetry points, these two pairs of states are inverted as shown in the schematic figure. For systems with three-fold rotational symmetry ($C_{3i}$), in the first phase (adiabatically connected with $H_1$), the parity Chern number is $Q_1=1$ and thus we expect one topological nodal line in the upper half of the Brillouin zone (and one in the lower half) for a generic Hamiltonian in this phase. As we deform the Hamiltonian to $H_1$, the nodal line shrink into  a Dirac point located on the $k_z$ axis. In the second phase,  we have $Q_2=3$ and thus are three topological nodal lines in the upper  (lower) half of the zone. As we deform this Hamiltonian into $H_2$, these nodal lines merge into three Dirac points, which are not located around the $k_z$ axis, in contrast to $H_1$. As we try to cross the phase boundary (say from the first phase to the second one), the two nodal lines in the first phase (one above and one below the $k_z=0$ plane) moves towards each other and eventually they meet at the $k_z=0$ plane. After the two nodal lines meet, they regroup and split into six new nodal lines (three above and three below $k_z=0$). The phase transition involves a gap closing in the $k_z=0$ plane (which was fully gapped in both phases away from the phase boundary), which enables the topological index to change its value, i.e. a topological phase transition.

In systems with four- or six-fold rotational symmetry (point group $C_{4h}$ or $C_{6h}$), the two quantum phases have $Q_1=2$ and $Q_2=0$, and thus topological nodal lines with nontrivial monopole charge (blue/yellow shown in the figure) arise in the first quantum phase but not in the second. In both phases, there is a mirror-symmetry protected nodal line (green) in the $k_z=0$ plane. At the topological phase transition, the two nodal lines with nontrivial monopole charge meet and annihilate at the $k_z=0$ plane.

It is worthwhile to emphasize that the topological nodal lines here each carry monopole charge $2$, instead of $1$. And thus, as this pair of nodal lines annihilate at the topological phase transition, the topological index changes by $2$ instead of $1$. This is in direct contrast with generic cases where the monopole charge can only be $0$ or $1$. Due to the four-fold (six-fold) rotational symmetry here, the topological index (monopole charge) is doubled. Instead of $0$ and $1$, it takes the value $0$ or $2$. In contrast to a generic nodal line with monopole charge $1$, a monopole charge $2$ nodal line can only arise in systems with four- or six-fold rotational symmetry. Furthermore, such a charge-$2$ topological nodal line must be located around a high symmetry axis (with four- or six- fold rotational symmetry), and the nodal line itself is invariant under four- or six- fold rotation, in direct contrast to a conventional charge-$1$ nodal line. Furthermore, as we deform the Hamiltonian towards $H_1$, this nodal line shrink into a nodal point. This nodal point is a quadratic band crossing point similar to the one mentioned above in family 4. This quadratic band crossing also indicates that the topological index is $2$ here, instead of $1$.

In family 6, we consider two pairs of type-b bands with different rotational symmetry (symmetry representation $\Gamma_m^+$, $\Gamma_n^+$ and $\Gamma_{m'}^-$, $\Gamma_{n'}^-$  with $m,n,m',n'$ being all distinct). This case can only arise in systems with six-fold symmetry ($C_{6h}$), because other rotational groups don't have two different pairs of type-b representations. In this case there is always a nodal point long the z-axis for $0<k_z<\pi$ (and another one for $-\pi<k_z<0$), because all bands have different rotational symmetries along the $k_z$ axis. This nodal point is a Dirac point and it carriers monopole charge $1$.

Again, we have two topologically-distinct quantum phases with $Q_1=3$ and $Q_2=1$ respectively. At the topological phase transition, one pair of nodal lines (one above $k_z=0$ and one below) meet and annihilate at the $k_z=0$ plane. This nodal line preserves the six-fold rotational symmetry, and it carriers monopole charge $2$, similar to the ones we observed in family 5 above. As we deform the Hamiltonian towards $H_1$, naively, we shall expect this nodal line to shrink into a quadratic band crossing point same as the monopole-charge-$2$ nodal line in family 5. However, because we have another Dirac point on the $k_z$ axis, this nodal line (with index 2) and the Dirac point (with index 1) will all merge together into one nodal point, which is a cubic band crossing point, consistent with the total topological index $Q_1=3$.

\section{Discussion}
\subsection{Higher symmetries}
Above, we considered the space-inversion symmetries and $n$-fold rotational along $z$, with $n=1$, $2$, $3$, $4$, $6$, which corresponds to point group symmetries $C_i$, $C_{2h}$, $C_{3i}$, $C_{4h}$, $C_{6h}$ respectively. For crystals with any of these point group symmetries, we have systematic demonstrated  the method to determine topological indices based on high-symmetry-point band inversions and discussed the implication of these topological indices in terms of topological nodal lines and nodal points.

For systems with more complicated point group symmetry, most of our techniques and conclusions can be generalized. This is because  as shown above, the index that we used doesn't  require any additional symmetry to be defined or calculated, and thus the topological nodal lines presented above do not require additional symmetries to remain stable. Therefore, if we are dealing with a crystal with higher point group symmetries, we can introduce a small perturbation to break the symmetry down to one of the groups that we considered here, and all our conclusions and nodal lines should remains stable as we turn off this small perturbation to recover the full symmetry of the crystal.

On the other hand, it must be emphasized that in systems with a higher symmetry, new phenomena and new topological nodal lines may arise. For example, if the point group gives us additional mirror planes, extra mirror-symmetry-protected nodal lines may emerge, which is beyond the scope of our study.

\subsection{higher monopole charge due to rotational symmetry}
As we know, generically, the monopole charge of a topological nodal line can only be $0$ or $1$, i.e. a $Z_2$ index. However, as mentioned above, for systems with four-fold or six-fold rotational symmetry, we observed nodal lines which carry higher monopole charge beyond $1$. This higher index is due to the rotational symmetry. Here, we use the $C_{4h}$ group in family 4 and 5 to demonstrate this high index, and to prove that our conclusion remains stable when extra bands are introduced (i.e. beyond four-band models). Instead of a four-band system, we consider a system with arbitrary number of bands. First, we will consider a simpler case where all the bands are type-a (like in family 4) or all of them are type-b (like in family 5). Later, we will discuss generic systems with  both type-a and type-b bands.

Here, we consider a $N$-band system composed of type-a orbits only, while the same conclusions hold for systems with only type-b orbits.
For a nodal line cage (family 4) or a monopole-charge-2 nodal line (family 5) shown in Fig.~\ref{table}, we can enclose them with a closed-2D manifold. In particular, we can choose the
manifold such that it is invariant under two-fold rotation ($C_2$) along the $k_z$ axis, which can always be achieved because the nodal lines or cages here must preserve this $C_2$ symmetry. For this closed manifold, we can define the monopole charge, and it has been shown that the monopole charge should be a $Z_2$ index, if no rotational symmetry is assumed. However, in this case that we consider here, as will be shown below, we can define a $Z_2$ monopole charge for half of the manifold (say the half with $0<k_x<\pi$). And due to the two-fold rotational symmetry ($C_2$), the other half of the manifold must carrier the same $Z_2$ monopole charge. As a result, the total monopole charge of the manifold can be either $0$ or $2$, depending on whether the half-manifold monopole charge is $0$ or $1$. This is the fundamental reason why we can obtain high monopole charge in this case, which results in a new type of topological nodal lines. To define the half manifold  index, we utilizes the two-fold rotational symmetry along $z$. It can be easily checked that if we have only type-a (or type-b) bands in our system, the $C_{4h}$ symmetry ensures that the two-fold rotation must be an identity operator ($+I$ if we only have type-a bands and $-I$ if we only have type-b bands). As a result,
\begin{align}
H_{C_2 \mathbf{k}}=C_2 H_{\mathbf{k}} C_2^{-1}= H_{\mathbf{k}}.
\end{align}
Here, we utilized the fact that $H$ and $C_2$ commute with each other, since $C_2$ is an identity operator (up to a -1 sign for type-b bands).
As a result, for the manifold we choose, an extra symmetry is obtained $H_{C_2 \mathbf{k}}= H_{\mathbf{k}}$. This symmetry implies that as far as our Hamiltonian is concerned, we can merge momentum points $C_2 \mathbf{k}$ and $\mathbf{k}$. For a half manifold  ($0<k_x<\pi$), this procedure sews together momentum points on the edge, and transforms the half manifold into a compact manifold. Thus, we can define a $Z_2$ monopole charge for it. And it is easy to verify that the other half manifold must have the same monopole charge.

In the proof above, we assumed that the system only contains type-a (or type-b) bands. In a real systems, in principle, both types of bands shall exist. In this part, we argue that our conclusion will remain as along as other bands does not come close to the Fermi energy. The argument is based on the stability of the topological nodal lines that we found in family 4 and 5 (with $C_{4h}$ symmetry) which carriers monopole charge $2$. As our proof above has showed, these monopole-charge-2 nodal lines remain stable if we go beyond four-band model, as long as the extra bands that we added to the system has the same time-reversal type. In this section, we will introduce extra bands with different time reversal type as a perturbation and see if these nodal lines remain stable. Without loss of generality, here again we start from the case with only type-a bands and a monopole-charge-2 nodal-line cage, and then introduce some extra type-b bands at energy away from the Fermi energy. For a topological nodal line, it is known that as long as the time-reversal and space-inversion symmetry is preserved (and spin-orbit couplings remain negligible), a nodal line cannot be gapped out unless it is first shrunk into a nodal point. As we discussed above, for a monopole-charge-2 nodal line, it has to preserve the rotational symmetry and thus when it shrinks into a nodal point, it must be a point along the high symmetry axis ($k_x=k_y=0$ in this case). Along such a high symmetry line, type-a and type-b bands cannot hybridize, because they have different rotational eigenvalues under a $C_4$ rotation. As a result, even if extra type-b bands are introduced, they are incapable of gapping out this nodal line/point. This argument shows that at least in the perturbative range, these high-monopole-charge nodal lines are stable, regardless of how many bands we include in the system and what type of bands we include.

What we will not consider here are extra type-b bands added near the Fermi energy (inside the gap). This would be a much more complicated case, and whether the interplay between type-a and type-b bands at the Fermi energy can generate interesting novel phenomena remains an open question and will be explored in a separate project.

The same analysis can be generalized to certain cases with six-fold rotations, which proves the stability of the nodal line with monopole charge $2$ in systems with $C_{6h}$ symmetry.

\acknowledgements{H.L. and K.S. acknowledge the support of the National Science Foundation Grant No. EFRI-1741618.}

\appendix
\section{Definition of the $\Gamma$ matrices}
In this section we write down the explicit form of $\Gamma$ matrices that we adopted. This set of matrices is not unique, and is sensitive to the choice of bases. For our convenience, we  will use the following basis.
\begin{align}
\Gamma_1=&\tau_z\otimes\sigma_0, \nonumber\\
\Gamma_2=&\tau_x\otimes\sigma_z,\nonumber\\
\Gamma_3=&\tau_x\otimes\sigma_x, \nonumber\\
\Gamma_4=&\tau_x\otimes\sigma_y,\nonumber\\
\Gamma_5=&\Gamma_1\Gamma_2\Gamma_3\Gamma_4=-\tau_y\otimes\sigma_0,  \nonumber\\
\end{align}
Here $\tau_i$ and $\sigma_i$ with $i=0,x,y,z$ are the $2\times 2$ identity and Pauli matrices where
\begin{align}
\tau_0=&\sigma_0=\left(
  \begin{array}{cc}
    1 & 0 \\
    0 & 1
  \end{array}
\right), \ \tau_x=\sigma_x=\left(
  \begin{array}{cc}
    0 & 1 \\
    1 & 0
  \end{array}
\right), \nonumber \\
 \tau_y=&\sigma_y=\left(
  \begin{array}{cc}
    0 & -i \\
    i & 0
  \end{array}
\right),\ \tau_z=\sigma_z=\left(
  \begin{array}{cc}
    1 & 0 \\
    0 & -1
  \end{array}
\right)
\end{align}

In the dual basis, we have
\begin{align}
\tilde{\Gamma}_1=&\tau_z\otimes\sigma_0, \nonumber\\
\tilde{\Gamma}_2=&\Sigma_{14}=-\tau_y\otimes\sigma_y,\nonumber\\
 \tilde{\Gamma}_3=&i \Gamma_2\Gamma_3\Gamma_4=-\tau_x\otimes\sigma_0,\nonumber\\
\tilde{\Gamma}_4=&-\Sigma_{12}=\tau_y\otimes\sigma_z,\nonumber\\
\tilde{\Gamma}_5=&\tau_y\otimes\sigma_x
\end{align}
In both these two bases, $TI=K$ with $K$ being the complex conjugation, although the explicit form of $T$ and $I$ are model-dependent and not universal.

\section{Reduced Hamiltonian}
Here we consider a four-band model as discussed in the main text, and prove that they can be reduced to Hamiltonian $H_1$ or $H_2$.

First, we start from a generic four-band model without assuming the time-reversal or space-inversion symmetry. As shown in the main text, we can write the Hamiltonian
in terms of the sixteen matrices that we choose, which are $\Gamma$ matrices and their products.
\begin{align}
H=a \Gamma_1+b_1 \Gamma_2+ b_2\Gamma_3 + c_1 \Sigma_{14}+c_2 i \Gamma_2\Gamma_3\Gamma_4 +\ldots
\label{eq_Ham_app}
\end{align}
where $a$, $b_1$, $b_2$, $\ldots$ are real functions of the momentum $\mathbf{k}$.
For a system made of four orbitals, where two bands constructed by the first two orbitals are inverted with the other two bands constructed by the last two orbitals,
one of these 16-matrices is special and we choose it to be the $\Gamma_1$ matrix
\begin{align}
 \Gamma_1=\left( \begin{array}{cccc}
    +1 &0 & 0 & 0\\
    0 & +1 & 0 & 0\\
    0 & 0 & -1 & 0\\
    0 & 0 & 0 & -1
  \end{array}
\right)
\end{align}
This matrix marks the band inversion. If we turn off all other terms, when the coefficient of this matrix [i.e. $a$ in Eq.~\eqref{eq_Ham_app}] is positive (negative) the first two
orbits have positive (negative) energy. Thus, the double band inversion that we consider in the manuscript is represented by the change of sign
in the coefficient $a$. For a small band inversion at $\Gamma$, $a$ is positive near the $\Gamma$ point, and it changes sign as we move away from $\Gamma$.
These two regions (with $a>0$ and $a<0$ respectively) are separated by a 2D manifold, on which $a=0$. For a small band inversion at $\Gamma$,
this 2D manifold has the topology of a sphere.

The eigenvalues of a Hamiltonian can be computed via the eigen-equation. $\det (H- \epsilon I)=0$. For a four-band model, this equation is a fourth-order equation of $\epsilon$.
Because the absolute value of energy plays little role in topology, we can make the Hamiltonian traceless by deducting its trace part, without changing any topological features.
For a traceless Hamiltonian, the eigen-equation must take the following structure
\begin{align}
\epsilon^4+q \epsilon^2+ r \epsilon + s=0
\end{align}
where $q$, $r$ and $s$ depend on the 16 coefficients in Eq.~\eqref{eq_Ham_app}. Here, the $\epsilon^3$ term is absent, because its coefficient is the trace of $H$,
which we set to zero.

For a system with two conduction and two valence bands, we can always make the two conduction bands degenerate, and same can be done to the valence bands. As long as
we don't introducing new band crossings in this procedure, the topological information remains. For nodal lines with nontrivial monopole charge, this procedure shrinks a nodal line into
a nodal point with four-fold degeneracy.

After this procedure, our fourth-order eigen-equation has two set of degenerate solutions, i.e. two conduction bands with identical eigen-energy $+\epsilon$ and two valence bands
with  $-\epsilon$. In other words, our eigen-equation shall have two pairs of equal real roots.
For a quartic equation shown above, this requires the following conditions: (a) the discriminant is zero
\begin{align}
\Delta=16 q^4 s-4q^3r^2-128 q^2 s^2+144 q r^2 s-27 r^4+256 s^3
=0
\end{align}
and (b) $q<0$ and (c) $q^2=4 s$. For a generic fourth order equation, these three conditions are independent. However, as an eigen-equation of a Hermitian matrix,
the last condition (c) is sufficient to guarantee (a) and (b).
This is because for an Hermitian matrix, all roots of the eigen-equation have to be real, and this constraint makes (c) sufficient to ensure the other two.

In general, the condition $q^2=4 s$ here is a very complicated equation. In terms of the $16$ coefficients in Eq.~\eqref{eq_Ham_app}, this equation is a
quartic equation for those coefficients and it involves nearly $200$ terms. However, luckily, such a complicated equation can be simplified in the form shown in Eq.~\eqref{eq:degenerate_condition}, i.e.,
the sum of 15 squares equals to 0. Because all the coefficients of Eq.~\eqref{eq_Ham_app} are real, sum of squares being zero means that every single square needs to
be zero, which gives us 15 quadratic  equations [Eq.~\eqref{eq:Delta_1}-\eqref{eq:Delta_15}].
These 15 equations are not independent. For the problem that we considers in the manuscript, due to the time-reversal and space-inversion symmetry,
the Hamiltonian can be made real [i.e. $b_1'=b_2'=c_1'=c_2'=e'=f'=0$ in Eq.~\eqref{eq:Ham_matrixform}]. For such a real Hamiltonian, it turns out that if we assume
that conduction and valance bands are both degenerate and $H$ is traceless (i.e. $q^2=4 s$), these 15 equations allow us to have
5 free parameters in Eq.~\eqref{eq_Ham_app}, which are the
5 terms that we shown in this equation. The rest of coefficients can all be computed from these 15 equations and thus they are not independent parameters anymore. This is the
reason why we selected to show these 5 terms in our Hamiltonian.

Furthermore, we can further reduce the number of free parameters here by 2, by turning off either $b_{1/2}$ or  $c_{1/2}$. For a double-band inversion that we consider here
in most part of the k-space,  we have $a\ne 0$, and it it is easy to check that here we can turn off all other coefficients in $H$, and the system will remain gapped and non-singular.
On the other hand, for the 2D manifold with $a=0$, only two options are available: (a) $b_1^2+ b_2^2 \ne 0$ while all other matrices have zero coefficients or (b)
$c_1^2+ c_2^2 \ne 0$ while all other matrices have zero coefficients. Except for these two options, the 15 equations we mentioned above will tell us that certain coefficients
must diverge, due to terms like $c_1/a$.

In summary, when we make the two conduction and two valence bands into degenerate pairs (without introducing new band crossings between conduction and valence bands),
we can turn off any terms except for $a \Gamma_1$ in our Hamiltonian as long as $a\ne 0$. If $a=0$, we have to choose from the two options shown above.
As a result, for a $\Gamma$ point band inversion, if the $a=0$ manifold follows the option (a), we can turn off all other terms except for $a$, $b_1$ and $b_2$ in the whole $k$-space without introducing any singularity or gap closing. On the other hand, if the $a=0$ manifold follows option (b), then we can turn off all terms except for $a$, $c_1$ and $c_2$.

In either case, by making the conduction/valance bands degenerate, we can turn off almost all the terms in Eq.~\eqref{eq_Ham_app} and only keep three of them, which are
$H_1$ or $H_2$.

\begin{widetext}
\section{Conditions for band degeneracy}
Consider a generic $4\times 4$ tracless Hamiltonian
\begin{align}
H=
\left( \begin{array}{cccc}
    a+d_1 & (e+e' i)+(f+ f' i) & -(c_2+i c_2')+(b_1+b_1' i) & (b_2+ b_2' i)+(c_1+c_1' i)\\
    (e-e' i)+(f- f' i) & a-d_1 & (b_2+ b_2' i)-(c_1+c_1' i) & -(c_2+i c_2')-(b_1+i b_1')\\
    -(c_2-i c_2')+(b_1-b_1' i) & (b_2- b_2' i)-(c_1-c_1' i) & -a+d_2 & (e+e' i)-(f+ f' i)\\
    (b_2- b_2' i)+(c_1-c_1' i) &  -(c_2-i c_2')-(b_1-i b_1') & (e-e' i)-(f- f' i) & -a-d_2
  \end{array}
\right)
\label{eq:Ham_matrixform}
\end{align}
the condition for degenerate bands $q^2=4 s$ can be written as
\begin{align}
0=q^2/4-s=\sum_{i=1}^{15} \Delta_i^2
\label{eq:degenerate_condition}
\end{align}
where
\begin{align}
\Delta_1&=a (d_1-d_2)-2 (b_1 c_2+ b_1' c_2')\label{eq:Delta_1}
\\
\Delta_2&=a (d_1+d_2)+2 (b_2 c_1+b_2' c_1')\\
\Delta_3&=2(c_2' c_1-c_1' c_2+a f')\\
\Delta_4&=2(-b_1 c_1-b_1' c_1'+a e)\\
\Delta_5&=2(-b_2 c_2-b_2' c_2'+a f)\\
\Delta_6&=2(b_2 b_1'-b_2' b_1+a e')\\
\Delta_7&=-2 e' c_2'+ 2f b_1- b_2(d_1-d_2)\\
\Delta_8&=-2 e' c_2- 2f b_1'+b_2'(d_1-d_2)\\
\Delta_9&=-2 e c_2+2 f' b_1'+c_1(d_1-d_2)\\
\Delta_{10}&=-2 e c_2'-2f' b_1+c_1'(d_1-d_2)\\\
\Delta_{11}&=2 e b_2'-2e' c_1+ b_1'(d_1+d_2)\\\
\Delta_{12}&=2 e b_2+2 e' c_1'+ b_1(d_1+d_2)\\
\Delta_{13}&=-2f c_1-2 f' b_2'-c_2(d_1+d_2)\\
\Delta_{14}&=-2f c_1'+2 f' b_2-c_2' (d_1+d_2)\\
\Delta_{15}&=2e f+2e' f'+(d_1^2-d_2^2)/2
\label{eq:Delta_15}
\end{align}
Because all the terms here are real, the sum of squares equals to zero implies $\Delta_i=0$ for any $i$.

\section{lattice models}
In this section, we show the lattice models that we used to generate the nodal-line figures in Fig.~\ref{table} in the main text. First we define some functions
for representations in different point groups for future convenience:
\bea
C_{3i}: \ \  \ \Gamma_1^+: \ \  f_1&=&\cos k_1+ \cos k_2+ \cos k_3  \nonumber\\
\Gamma_1^+: \ \  f_2&=&\cos k'_1+ \cos k'_2+ \cos k'_3, \  \nonumber\\
\Gamma_2^-: \ \  f_3&=&\sin (k_1+k_z)+\ w^2 \sin(k_3+k_z)\ -w \sin(-k_2+k_z)\nonumber\\
\Gamma_2^-: \ \ f_4&=&\sin (k'_1+k_z)+\ w^2 \sin(k'_3+k_z)\ -w \sin(-k'_2+k_z)\nonumber\\
\Gamma_1^-: \ \ f_5&=&\sin (k_1+k_z)+\ \sin(k_3+k_z) + \sin(-k_2+k_z)\nonumber\\
\Gamma_1^-: \ \ f_6&=&\sin (k'_1+k_z)+\ \sin(k'_3+k_z) +\ \sin(-k'_2+k_z) \nonumber\\
C_{6h}: \ \ \  \Gamma_2^-: \ \  f_7&=&(\cos k_1 - w_6^2 \cos k_2 +w_6^4 \cos k_3) \sin k_z\nonumber\\
\Gamma_4^-: \ \ f_8&=&\sin k_1 -\sin k_2+\sin k_3  \nonumber\\
\Gamma_4^-: \ \ f_9&=&\sin k'_1 -\sin k'_2+\sin k'_3 \nonumber\\
\Gamma_4^+: \ \ f_{10}&=&(\sin k_1 -\sin k_2+\sin k_3) \sin k_z \nonumber\\
\Gamma_5^-: \ \ f_{11}&=&\sin k_1+w_6^2\sin k_2+w_6^4\sin k_3
\eea
\bea
k_1&=& k_x, \ k_2=\frac{1}{2}k_x+\frac{\sqrt{3}}{2}k_y, \ k_3=-\frac{1}{2}k_x+\frac{\sqrt{3}}{2}k_y \nonumber\\
k'_1&=& \frac{3}{2}k_x+\frac{\sqrt{3}}{2}k_y, \ k'_2=\sqrt{3}k_y, \ k'_3=-\frac{3}{2}k_x+\frac{\sqrt{3}}{2}k_y \nonumber\\
w&=&e^\frac{i\pi}{3}, \ \ \  w_6=e^\frac{i\pi}{6}
\eea
Here the functions defined with $k_{1,2,3}$ are for nearest-neighbour hoppings, and the functions with $k'_{1,2,3}$ are for next-nearest neighbour hoppings.

\subsection{Family 1, $C_{i}$ and $C_{3i}$}
For systems with $C_{i}$ ($C_{3i}$) symmetry, we consider a triclinic (hexagonal) lattice with four orbits per site and the symmetry of these four orbits are $\Gamma_1^+$, $\Gamma_1^+$, $\Gamma_1^-$ and $\Gamma_1^- $. Using these four orbits as the basis for our Hamiltonian, we have $T=K$ and $I=\tau_z\otimes \sigma_0$ and the Hamiltonian is
\bea
H&=&a\ \tau_z \otimes \sigma_0+b_1 \ \tau_y\otimes \sigma_z+b_2\ \tau_y\otimes\sigma_x-c_1\ \tau_x\otimes\sigma_y
\eea

For the system with the  $C_i$ symmetry shown in Fig.~\ref{table},  the parameters  are:
\bea
a&=& 1.5(\cos k_x+1.2 \cos k_y) +1.5 \cos k_z -2.5 \nonumber\\
b_1&=& \sin k_y+0.5 \sin k_z \nonumber\\
b_2&=& \sin k_x-\sin k_y \nonumber\\
c_1&=& 0.3 (\sin k_z+0.5 \sin k_y)
\eea
Here, only nearest-neighbor hoppings are needed, and the longer-range hoppings are set to zero.

The parameters for the $C_{3i}$ system are:
\bea
a&=&f_1+f_2+4 \cos k_z-6 \nonumber\\
b_1&=&0.2 f_6 \nonumber\\
b_2&=&0.4 f_5 \nonumber\\
c_1&=&0.7 \sin k_z
\eea
where $f_2$ and $f_6$ are from next-nearest neighbor hoppings and all others are from nearest neighbor hoppings.

\subsection{Family 4, $C_{2h},\ C_{4h},\ C_{6h}$}
Here, same as before, we put four orbits on each lattice site with symmetry $\Gamma_1^+$, $\Gamma_2^+$, $\Gamma_1^-$ and $\Gamma_2^-$ for $C_{2h}$ and $C_{4h}$, and
$\Gamma_1^+$, $\Gamma_4^+$, $\Gamma_1^-$, $\Gamma_4^-$ for $C_{6h}$. Using these four orbits as the basis for our Hamiltonian, we have $T=K$,
$P=\tau_z\otimes \sigma_0$. The two-fold rotation is $C_2=\tau_0\otimes \sigma_z$ for $C_{2h}$, while the four-fold rotation is $C_4=\tau_0\otimes \sigma_z$ for $C_{4h}$
and $C_6=\tau_0\otimes \sigma_z$ for $C_{6h}$.

The Hamiltonian is
\bea
H&=&a\ \tau_0 \otimes \sigma_z +b_1 \ \tau_0\otimes \sigma_x-b_2\ \tau_x\otimes\sigma_y+g\ \tau_z\otimes\ \sigma_0
\eea

The parameters for a system with $C_{2h}$ symmetry are
\bea
a&=&\cos _x+\cos k_y+1.5 \cos k_z -2 \nonumber\\
b_1&=&\sin k_x \sin k_z \nonumber\\
b_2&=&\sin k_y  \nonumber\\
g&=&0.3
\eea
Here, only nearest-neighbor hoppings are needed, and the longer-range hoppings are set to zero.

The parameters for the nodal lines in $C_{4h}$ are:
\bea
a&=&\cos _x+\cos k_y+1.5 \cos k_z -2 \nonumber\\
b_1&=&\cos k_x -\cos k_y \nonumber\\
b_2&=&\sin k_z (\cos(k_x+k_y)-\cos(k_x-k_y))  \nonumber\\
g&=&0.2
\eea
The $b_2$ above is from next-nearest neighbor hopping while all the other terms are from nearest neighbor hopping.

The parameters for the nodal lines in $C_{6h}$ are:
\bea
a&=& f_1+f_2+4\cos k_z -6 \nonumber\\
b_1&=&5f_{10} \nonumber\\
b_2&=&f_9  \nonumber\\
g&=&0.4
\eea
where $f_2$ and $f_9$ are from next-nearest neighbor hoppings and all other terms are from nearest neighbor hoppings.

\subsection{Family 5, $C_{3i}$}
Here, we put four orbits with symmetry $\Gamma_2^+$, $\Gamma_3^+$, $\Gamma_2^-$, $\Gamma_3^-$ on each lattice site, and use these four orbits as the basis of
the Hamiltonian. As a result, we have $T=\tau_0\otimes \sigma_x K$, $P=\tau_z \otimes \sigma_0$. Rotation operators are given by the representations $C_3=\textrm{diag}(\omega^2,-\omega,\omega^2,-\omega)$, where $\omega=e^{i\frac{\pi}{3}}$ and diag refers to a diagonal matrix. The Hamiltonian for the nodal lines is:
\bea
H&=&a\ \tau_z \otimes \sigma_0+b_1 \ \tau_y\otimes \sigma_y+b_2\ \tau_y\otimes\sigma_x+c_1\ \tau_x\otimes\sigma_z+c_2\ \tau_y\otimes \sigma_0  \nonumber\\
a&=&f_1+0.3f_2+4 \cos k_z-4 \nonumber\\
b_1+ib_2&=&(1-g_0) (0.5f_3+0.3 f_4) \nonumber\\
c_1&=&0.4g_0 f_5 \nonumber\\
c_2&=&0.2g_0 f_6
\eea
The figure with $Q_1=1$ has $g_0=0.4$ and the one with $Q_2=3$ has $g_0=0.8$. The terms $f_2$, $f_4$ and $f_6$ are from next-nearest neighbor hoppings and the rests are from nearest neighbor.

\subsection{Family 5, $C_{4h}$}
Here, the four orbits have symmetry  $\Gamma_3^+$, $\Gamma_4^+$, $\Gamma_3^-$, $\Gamma_4^-$ and thus $C_4=\textrm{diag}(i,-i,i,-i)$, $T=\tau_0\otimes\sigma_x K$, and $P=\tau_z\otimes \sigma_0$. The Hamiltonian is
\bea
H&=&a\ \tau_z \otimes \sigma_0+b_1 \ \tau_y\otimes \sigma_y+b_2\ \tau_y\otimes\sigma_x+c_1\ \tau_x\otimes\sigma_z \nonumber\\
a&=& \cos k_x+\cos k_y +1.5 \cos k_z -2 \nonumber\\
b_1&=& (1-g_0)\left( \sin k_z (\cos k_x - \cos k_y)+0.2 \sin k_z (\cos (k_x+k_y)-\cos(k_x-k_y)) \right) \nonumber\\
b_2&=& (1-g_0)\left(0.2 \sin k_z (\cos k_x - \cos k_y)+ \sin k_z (\cos (k_x+k_y)-\cos(k_x-k_y)) \right) \nonumber\\
c_1&=& g_0 \sin k_z
\eea
The figure with $Q_1=2$ has $g_0=0.25$ and the one with $Q_2=0$ has $g_0=0.7$. As $g_0$ becomes larger, the two nodal lines with nontrivial monopole charge move towards the $k_z=0$ plane and annihilate with each other. Terms proportional to $(\cos k_x - \cos k_y)$ are from nearest neighbor hoppings and those proportional to $(\cos (k_x+k_y)-\cos(k_x-k_y))$ are from next-nearest neighbor hoppings.

\subsection{Family 5, $C_{6h}$}
Here, the four orbits have symmetry $\Gamma_2^+$, $\Gamma_3^+$, $\Gamma_2^-$, $\Gamma_3^-$, and $C_6=\textrm{diag}(-w_6^2,w_6^4,,-w_6^2,w_6^4)$ with $w_6=e^{\frac{i\pi}{6}}$. $T=\tau_0\otimes\sigma_x K$, $P=\tau_z\otimes \sigma_0$. The Hamiltonian is
\bea
H&=&a\ \tau_z \otimes \sigma_0+b_1 \ \tau_y\otimes \sigma_y+b_2\ \tau_y\otimes\sigma_x+c_1\ \tau_x\otimes\sigma_z \nonumber\\
a&=& f_1+f_2+4 \cos k_z-6 \nonumber\\
b_1+ib_2&=& (1-g_0)f_7  \nonumber\\
c_1&=& g_0 \sin k_z
\eea
where $f_2$ is from next-nearest neighbor hopping and all other terms are from nearest neighbor hoppings. The figure with $Q_1=2$ has $g_0=0.04$ and the one with $Q_2=0$ has $g_0=0.6$. The two extra-robust nodal lines annihilate as $g_0$ grows.

\subsection{Family 6, $C_{6h}$}
Here, the four orbits have symmetry $\Gamma_2^+$, $\Gamma_3^+$, $\Gamma_5^-$, $\Gamma_6^-$. $C_6=\textrm{diag}(-w_6^2,w_6^4,,w_6^2,-w_6^4)$. The Hamiltonian is:
\bea
H&=&a\ \tau_z \otimes \sigma_0+b_1 \ \tau_y\otimes \sigma_y+b_2\ \tau_y\otimes\sigma_x+c_1\ \tau_x\otimes\sigma_z+c_2\ \tau_y\otimes \sigma_0,  \nonumber\\
a&=& f_1+f_2+4 \cos k_z-6 \nonumber\\
b_1+ib_2&=& (1-g_0)f_{11}  \nonumber\\
c_1&=& g_0 f_8 \nonumber\\
c_2&=& g_0 f_9
\eea
Here, $f_2$ and $f_9$ are from next-nearest neighbor hoppings, while all other terms are from nearest neighbor hoppings. The figure with $Q_1=3$ has $g_0=0.9$ and the one with $Q_2=1$ has $g_0=0.5$.
\end{widetext}


%

\end{document}